\documentclass[%
superscriptaddress,
reprint,
 amsmath,amssymb,
 aps,
 pre,
floatfix,
hyperref
]{revtex4-2}

\usepackage{csquotes}
\usepackage{graphicx}
\usepackage{dcolumn}
\usepackage{bm}
\usepackage[utf8x]{inputenc}
\usepackage[usenames,dvipsnames,svgnames,table]{xcolor}
\usepackage{selinput}
\usepackage{soul}
\usepackage[normalem]{ulem}
\usepackage{cancel}

\definecolor{cream}{RGB}{222,217,201}
\definecolor{kjpurple}{RGB}{142,86,173}
\definecolor{kjblue}{RGB}{36,113,163} 
\definecolor{apgreen}{RGB}{22,160,133}

\usepackage[normalem]{ulem}

\begin{document}

\title{The role of substrate mechanics in osmotic biofilm spreading}

\author{Anthony Pietz}
\affiliation{Institut für Theoretische Physik, Universität Münster, Wilhelm-Klemm-Str.\ 9, 48149 Münster, Germany} 

\author{Karin John}
\affiliation{Université Grenoble Alpes, CNRS, LIPhy, 38000 Grenoble, France}
\email{karin.john@univ-grenoble-alpes.fr}

\author{Uwe Thiele}
\affiliation{Institut für Theoretische Physik, Universität Münster, Wilhelm-Klemm-Str.\ 9, 48149 Münster, Germany} 
\affiliation{Center for Nonlinear Science (CeNoS), Universit\"at M\"unster, Corrensstr.\ 2, 48149 M\"unster, Germany}
\email[For correspondends:]{u.thiele@uni-muenster.de}
\homepage{https://www.uni-muenster.de/Physik.TP/~thiele/}
 
\begin{abstract}
Bacteria invade surfaces by forming dense colonies encased in a polymer matrix. Successful settlement of founder bacteria, early microcolony development and later macroscopic spreading of these biofilms on surfaces rely on complex physical mechanisms. Recent data show that on soft hydrogels, substrate rigidity is an important determinant for biofilm initiation and spreading, through mostly unknown mechanisms. 
Using a thermodynamically consistent thin-film approach for suspensions on soft elastic surfaces supplemented with biomass production we investigate \textsl{in silico} the role of substrate softness in the osmotic spreading of biofilms. We show that on soft substrates with an imposed osmotic pressure spreading is considerably slowed down and may be completely halted depending on the biomass production rate. 
We find, that the critical slowing down of biofilm spreading on soft surfaces is caused by a reduced osmotic influx of solvent into the biofilm at the edges, which results from the thermodynamic coupling between substrate deformation and interfacial forces. 
By linking substrate osmotic pressure and mechanical softness through scaling laws, our simple model semi-quantitatively captures a range of experimentally observed biofilm spreading dynamics on hydrogels with different architectures, underscoring the importance of inherent substrate properties in the spreading process.
\end{abstract}

\maketitle


\section{Introduction}

Bacteria spend most of their life attached to surfaces in structured colonies encased in a self-produced polymeric matrix, called biofilms, which are the prevalent form of life on earth \cite{Flemming2016}.  
 The organisation in biofilms confers bacteria a selective advantage over the individual, e.g., by increasing resistance to mechanical damage and antibiotic agents. 
Biofilm formation requires bacteria to transition from a free-swimming, individual lifestyle, to a sessile cooperative one. 
Since the discovery of the multicellular tissue-like behaviour of bacteria about 100 years ago, research has focused on dissecting the environmental cues and biological pathways driving bacterial adhesion and biofilm formation: genetic changes, intracellular signalling, and cell-cell communication \cite{Kazmierczak2015}.
However, bacteria in their natural environments are continuously exposed to physical and physico-chemical forces, including mechanical stresses \cite{Persat2015,Yan2019,Cont2020,dufrene2020mechanomicrobiology,Asp2022,Geisel2022}, osmotic pressure gradients \cite{Rubinstein2012,Seminara2012,Yan2017,zhang2014nutrient,zhang2015surface,Srinivasan2019,Paul2019,Kumar2024}, capillary \cite{Fauvart2012,de2015role,Trinschek2018} and wetting forces \cite{Trinschek2017, Kampouraki2022} and it remains often unclear how individuals or colonies integrate such physical cues. 
However, it is becoming increasingly known that many bacterial communities control and use the mechanical and physico-chemical properties of their surrounding habitat to maximise their chances of survival and dissemination \cite{Maier2021}. It is now well understood  that auto-produced matrix molecules act as osmolytes for biofilms grown on hydrogel substrates. The resulting flux of water with nutrients from the substrate into the biofilm can act as a driving force for lateral biofilm expansion.
It has also been demonstrated that capillary, wetting and adhesion forces play a major role in the dynamics of the advancing edge of biofilms \cite{Trinschek2017,Si2018,tam2022thin}, and swarming colonies  \cite{Fauvart2012,de2015role,Trinschek2018,kotian2020active,Ma2021}. Bacteria-produced surfactants allow the biofilm to overcome wetting-induced stalling of the biofilm edge \cite{Trinschek2017,Kumar2024a} and accelerate spreading in swarming colonies through Marangoni-flows, that may also trigger a fingering instability at the spreading front \cite{Trinschek2018,Kumar2024a}.

The influence of other material properties on biofilm spreading is less well explored. An example is the role of the mechanical substrate stiffness that represents an important parameter for the growth of biofilms on and in soft tissues. Over the past decade, only a few studies have specifically focused on the role of substrate rigidity on bacterial dynamics, both at the single cell and community level  \cite{Saha2013,Song2014,Guegan2014,Siryaporn2014,Zhang2014,Arias2020a,Cont2020,Blacutt2021,Asp2022,Gomez2023,Ziege2021,Koch2022,Rashtchi2024,Faiza2024,Techakasikornpanich2024}. 
Their results show that the rigidity of the underlying substrate does impact bacterial attachment and motility as well as colony morphology and dynamics. However, the observed trends are not consistent between different bacterial strains (e.g. \textit{Pseudomonas aeruginosa}, \textit{Escherichia coli}), different types of soft substrates (hydrogels, elastomers, layer-by-layer polymeric substrates) and different explored stiffness ranges (kPa-MPa). 
Thereby, one of the main experimental difficulties is the independent variation of physico-chemical properties (e.g., osmotic pressure) and mechanical rigidity of the substrate, which might explain opposing trends observed on different hydrogel architectures with identical mechanical properties (shear modulus). 
Within this context, the osmotic spreading of biofilms has been studied on various substrate architectures and chemical compositions \cite{Asp2022}. There it has been shown that at constant osmotic pressure, spreading is slower on soft than on rigid substrates, raising the question about the underlying coupling mechanism between the dynamics of the biofilm edge and the substrate rigidity.

Motivated by the aforementioned observations, here, we numerically study the effect of the rigidity of a soft elastic substrate on osmotic biofilm spreading. To that end we employ a minimal thin-film model which naturally captures the effect of interfacial forces and elastic substrate deformations on the motion of the advancing biofilm edge. Briefly, we consider the biofilm as a shallow drop of viscous suspension whose lateral spreading is driven by bioactive growth and ensuing osmotic fluxes. The  relevant evolution equations for the biofilm components and the substrate deformation are derived from a gradient dynamics approach \cite{Thie2018csa,Henkel2021} supplemented by a bioactive growth term that mimics biological growth processes (cell division and matrix production). 

Within this framework we investigate how substrate rigidity and biomass production rate affect the lateral spreading speed at the biofilm edge. We recover the experimentally observed behaviour of a slowing down of spreading with increasing substrate softness, i.e.\ with decreasing rigidity, at constant substrate osmotic pressure. Furthermore, we identify a growth regime, where the substrate softness leads to a complete arrest of the biofilm edge. We find, that the reduced biofilm spreading speed on soft surfaces is not directly related to the visco-elastic braking observed for the spreading of droplets of passive liquid on soft solid surfaces \cite{Karpitschka2015,Carre2001,Andreotti2020,Henkel2021} but is rather caused by a reduced osmotic influx of solvent into the biofilm at its edges, which results from the thermodynamic coupling between substrate deformation and interfacial forces.

The manuscript is structured as follows. In section~\ref{sec:model} we introduce a minimal dynamic model for osmotic spreading on a soft elastic substrate and discuss the relevant parameters. In sections~\ref{sec:passive-results}-\ref{sec:comp} the model behaviour is analysed using full numerical simulations. First, in section~\ref{sec:passive-results} we investigate equilibrium biofilm shapes and substrate deformations for model biofilms with an imposed amount of biomass.
Then, in section~\ref{sec:active-results} we analyse the dependence of the spreading dynamics of active model biofilms on the substrate softness and biomass growth rate. Finally, in section~\ref{sec:comp}, we establish a link between our theoretical results and the experiments of ~\citet{Asp2022}, and thereby propose a unifying description of the qualitatively different dependencies of the spreading velocity on the substrates stiffness on hydrogels of various architectures. We conclude in section~\ref{sec:concl} by a thorough discussion of our theoretical results.

\begin{figure}[hbt]
\includegraphics[width=\hsize]{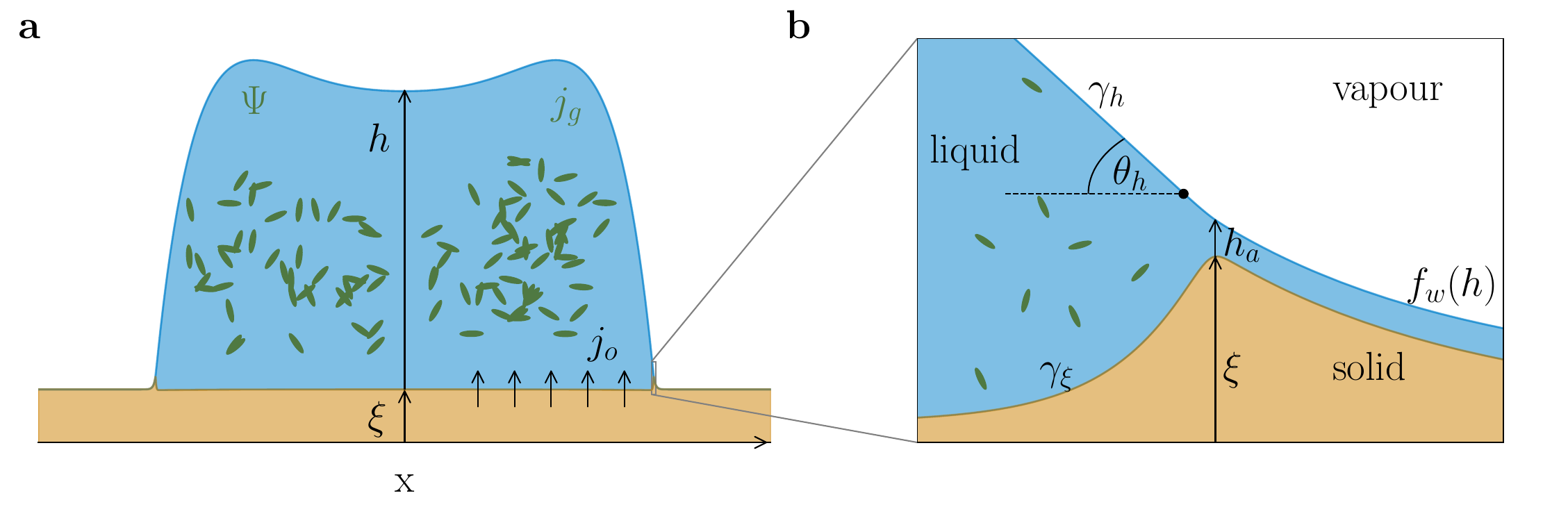}
\caption{Osmotic spreading of a biofilm on a soft substrate. (a) Sketch of a biofilm illustrating the film height $h(\vec{r},t)$, biomass amount $\Psi(\vec{r},t)$ and vertical substrate displacement $\xi(\vec{r},t)$. Osmotic pressure gradients are generated as bacteria consume water and nutrients to produce biomass via bacterial proliferation and matrix secretion, which is described by the growth term $j_g$.This causes an osmotic influx of nutrient-rich water $j_o$ from the moist substrate into the biofilm. (b) Magnification showing the mesoscopic view of the leading edge of the biofilm, illustrating the physical influences (capillarity, wettability, elasticity) in the contact line region: a fluid adsorption layer of height $h_a$ and devoid of biomass is in contact with the biofilm with  the macroscopic contact angle $\theta_h$. The vertical force balance at the contact line deforms the substrate (displacement $\xi$). $\gamma_h$ and $\gamma_\xi$ denote the surface tensions of the biofilm-gas and the substrate-biofilm interfaces, respectively. The wetting energy $f_\mathrm{w}(h)$ describes the wettability of the substrate.}
\label{sketch}
\end{figure}

\section{Model}
\label{sec:model}

We consider osmotic biofilm growth and spreading on a soft solid substrate within a mesoscopic modelling framework (Fig.~\ref{sketch}) as previously introduced for rigid solid substrates \cite{Trinschek2016,Trinschek2017}. Here, we incorporate an underlying viscoelastic substrate in analogy to the approach introduced by Henkel \textit{et al.}\cite{Henkel2021,Henkel2022}, namely, by considering a strain energy related to the vertical displacement of the solid-liquid interface in the fully compressible case. 
The biofilm itself is modelled as a  thin liquid film of a suspension of biomass (bacteria and extracellular matrix) in a solvent (nutrient-rich water). The relevant field variables are the film thickness $h(\vec{r},t)$, the biomass $\Psi(\vec{r},t)$ and the vertical substrate displacement $\xi(\vec{r},t)$ where $\vec{r}=(x,y)^T$ are planar Cartesian coordinates. 
The biomass variable $\Psi$ represents the effective biomass height, i.e.\ any spatial variations in a vertical direction of the biofilm are neglected. In addition, for convenience, we introduce the biomass volume fraction $\phi$ which is related to the effective biomass thickness $\Psi$ by 
\begin{equation}
\phi=\frac{\Psi}{h} \,.\label{eq:phi}
\end{equation}
For a thorough discussion of the usage of $\Psi$  and $\phi$ see \citet{Trinschek2016}.
The free energy functional $\bar{F}$ (expressed in $h$, $\xi$, and $\phi$) that determines the physico-chemical driving forces for all transport processes for the passive (non-bioactive) suspension is
\begin{align}
\bar{F}[h,\xi,\phi]  = \int_{\Omega} \left[f_\mathrm{cap}(h,\xi)+ f_\mathrm{el}(\xi) + f_\mathrm{w}(h) + h f_\mathrm{m}\left(h,\phi\right) \right] \text{d}\Omega \label{free_energy_functional} 
\end{align} 
with the capillary energy  (liquid-vapour surface tension $\gamma_h$, solid-liquid surface tension $\gamma_\xi$)
\begin{align}
f_{\text{cap}} &= \frac{\gamma_h}{2} |\nabla(\xi + h)|^2 + \frac{\gamma_{\xi}}{2} |\nabla \xi|^2 \,, \label{cap}
\end{align}
where $\nabla=(\partial/\partial_x,\partial/\partial_y)^T$ denotes the planar gradient operator.
The elastic energy due to a vertical deformation of the substrate is modelled by a Winkler foundation model for a fully compressible substrate \cite{Henkel2021}
\begin{align}
f_{\text{el}} &= \frac{\kappa_v}{2} \xi^2\label{el} 
\end{align}
where $\kappa_v$ is an elastic constant. We employ a simple wetting energy for a partially wetting liquid consisting of the sum of a short-range stabilising and a long-range destabilising Van-der-Waals interaction \cite{bonn2009wetting, kalliadasis2007thin, popescu2012precursor} 
\begin{align}
f_\mathrm{w} = A \biggl( -\frac{1}{2h^2} + \frac{h_a^3}{5h^5} \biggr)\label{wet} 
\end{align}
where $A$ is a Hamaker constant and $h_a$ denotes the equilibrium thickness of a very thin adsorption layer at the substrate. The same form of the wetting energy has recently been employed in other thin-film models \cite{Thie2018csa,Henkel2021,John2010,Trinschek2017,Trinschek2018,stegemerten2022symmetry,VoTh2024jem}.

The entropy of mixing in the bulk of the liquid film is given by
\begin{align}
f_\mathrm{m} = \frac{k_B T}{a^3}\biggl[\phi \ln{\phi} + (1-\phi) \ln{(1-\phi)} \biggr] \,,\label{mix} 
\end{align}
where $a$ denotes an effective length scale characterising the biomass. $k_B T$ denotes the thermal energy.
Using the gradient dynamics approach advanced and reviewed in \cite{thiele2011note,thiele2013gradient,Xu2015,Thie2018csa,Henkel2021,HDGT2024l,VoTh2024jem} we have derived the transport equations of the passive mixture supplemented by a bioactive growth term $j_\text{g}(h)$, which breaks the variational structure and renders the suspension bioactive. 
The resulting evolution equations for the thermodynamic variables $h$, $\Psi$, $\xi$ are given by
\begin{align}
		\label{gradient_dynamics_h_t} \partial_{t} h &= \nabla \cdot 
		\biggl[ Q_{hh}    \nabla \frac{\delta F}{\delta h} +
		Q_{h\Psi} \nabla \frac{\delta F}{\delta \Psi}\biggr] + j_o \\
		\partial_{t} \Psi &= \nabla \cdot 
		\label{gradient_dynamics_Psi_t} \biggl[ Q_{ \Psi h}   \nabla \frac{\delta F}{\delta h} +
		Q_{\Psi \Psi} \nabla \frac{\delta F}{\delta\Psi}\biggr] + j_g	\\
		\label{gradient_dynamics_xi_t} \partial_{t} \xi &=	- \frac{1}{\zeta} \frac{\delta F}{\delta \xi}
\end{align} 
Note, that here we employ the variation of the free energy $F[h,\xi,\Psi]$ which relates to the free energy $\bar{F}[h,\xi,\phi]$ in \eqref{free_energy_functional} as $\bar{F}[h,\xi,\phi]=\bar{F}[h,\xi,\Psi/h]=F[h,\xi,\Psi]$. In the following $\phi$ should only be seen as a convenient short-hand notation for $\Psi/h$.
The mobilities $Q_{ij}$ form the symmetric and positive definite mobility matrix 
\begin{align}
\begin{pmatrix}
			Q_{hh} & Q_{h\Psi} \\
			Q_{\Psi h} & Q_{\Psi\Psi} \\
\end{pmatrix} &= 
	\frac{1}{3\eta}
	\begin{pmatrix}
		h^3 & h^2 \Psi \\
		h^2 \Psi & h \Psi^2 \\
	\end{pmatrix} + 
	\begin{pmatrix}
		0 & 0 \\
		0 & D\Psi \\
	\end{pmatrix}\,.
\end{align}
Here, $\eta$ denotes the composition-dependent biofilm viscosity
\begin{align}
	\eta = \eta_0[(1 - \phi) + \mu \phi ]\,,
\end{align}
and $\mu$ represents the ratio $\mu=\frac{\eta_b}{\eta_0}$ of the viscosities of the pure biomass $\eta_b$ and the pure solvent $\eta_0$. Further, 
$D=a^2/(6\pi\eta)$ denotes the biomass diffusivity consistent with the diffusion constant $D_\mathrm{diff}=k_BT/(6\pi\eta a)$ with $a$ being the typical biomass length scale as introduced in Eq.~\eqref{mix}.
Finally, $\zeta$ denotes the viscosity of the solid viscoelastic substrate. 

The non-conserved fluxes $j_o$ and $j_g$ correspond to the osmotic flux of solvent between the moist substrate and the biofilm and the active growth of  biomass in the biofilm, respectively.
Thereby, the osmotic flux is thermodynamically driven by the difference between the osmotic pressures in the biofilm $\Pi$ and in the substrate $\Pi_s$, i.e.\
\begin{equation}
j_o=Q_o\left(\Pi-\Pi_s \right), \label{eq:osm:flux}
\end{equation}
where the former is given as variation of the energy functional 
\begin{equation}
\Pi=-{\delta F\over \delta h} \label{def:Pi}
\end{equation}
and the latter is imposed as
\begin{equation}
  \Pi_s=-\frac{k_BT}{a^3}\ln{(1-\phi_\text{eq})}. 
  \label{eq:Pi_s}
\end{equation}
Eq.~\eqref{eq:Pi_s} implies that a thick biofilm of uniform height $h\gg h_a$ and biomass concentration $\phi_\text{eq}$ is in thermodynamic equilibrium with the substrate.
Note, that we treat the substrate as an infinite solvent reservoir at fixed uniform osmotic pressure $\Pi_s$ (i.e.\ a chemostat). In other words, we neglect any effect a substrate deformation might have on $\Pi_s$.
Finally, $Q_o$ denotes a mobility.

The active biomass growth $j_g$ is modelled as logistic growth with limited resources 
\begin{align}
j_g(h,\phi) = g \color{black} h \phi(1-\phi)\left( 1 - \frac{h\phi}{h^\ast\phi_\text{eq}} \right)f_\text{mod}(h,\phi) \label{eq:growth-term}
\end{align}
where $g$ is a rate constant, $\Psi^\ast=\phi_\text{eq}h^\ast$ gives the upper limit for the biomass layer thickness that can be sustained on a substrate, i.e.\ neglecting spatial gradients, the biofilm will reach a stationary state of uniform height $h^\ast$ and biomass concentration $\phi_\text{eq}$. Thereby, $\Psi^\ast$ is related to the thickness for which nutrient diffusion and consumption of nutrients by the bacteria throughout the vertical profile of the film equilibrate \cite{Ward2011,zhang2014nutrient}. 
The function $f_\text{mod}$ which governs the onset of biomass growth is given by
\begin{align}
f_{\text{mod}}(h,\phi) = \left( 1 - \frac{h_u\phi_\text{eq}}{h\phi} \right)\left[ 1 - \exp\left( \frac{h_a\phi_\text{eq}- h\phi}{h_a} \right) \right] \label{eq:fmod} 
\end{align}
and ensures that growth only starts if the biomass $\Psi$ exceeds a (small) critical value $\Psi_u=\phi_\text{eq} h_u$, such that no biomass is produced spontaneously in the adsorption layer of thickness $h_a<h_u$, and the biomass in the adsorption layer equilibrates at a small value $\Psi_a=h_a\phi_\text{eq}$. For effective biomass thickness $\Psi\gg \Psi_u$ the function $f_\text{mod} = 1$ and does not affect the biomass growth.

A core feature of the employed simple gradient dynamics-based modelling for thin films of mixtures is the approximation that vertical concentration gradients are small \cite{Trinschek2016,Thie2018csa}. Note that another class of models takes vertical concentration gradients into account \cite{Ward2011,tam2022thin}. However, \citet{tam2022thin} also show that such vertical dependencies are small for biologically realistic parameters further confirming our chosen modelling approach.

To facilitate the model analysis we introduce the vertical length scale
\begin{equation}
\mathcal{H}=h_a\,,\label{eq:H}
\end{equation}
the horizontal length scale 
\begin{equation}
\mathcal{L}=\sqrt{\frac{\gamma_h}{A}} h_a^2 \,,\label{eq:L}
\end{equation} 
and the time scale 
\begin{equation}
\mathcal{T}=3\frac{\eta_0\gamma_h h_a^5}{A^2} \,.\label{eq:T} 
\end{equation}
The scaling of space and time results in a nondimensional parameter, the substrate softness $s\sim 1/\kappa_v$ (i.e.\ the inverse of the substrate rigidity)
\begin{equation}
s={\mathcal{L}^2_{ec} \over  \mathcal{L}^2} \label{def:s}
\end{equation}
where the elastocapillary length $\mathcal{L}_{ec}=\sqrt{\gamma_h\over \kappa_v}$ enters naturally, which characterises the typical elastic deformation induced by surface tension. 

In the remainder of the manuscript all dimensional quantities will be expressed tacitly in the scales $\mathcal{H}$,  $\mathcal{L}$ and $\mathcal{T}$, e.g.~the biomass growth rate constant $g$ in units of $1/\mathcal{T}$, the biofilm height $h$ and biomass layer thickness $\Psi$ in units of $\mathcal{H}$, the horizontal distance $x$ in units of $\mathcal{L}$, etc. The nondimensional model equations and nondimensional parameters are summarised in Table \ref{tab::listed-all-parameter} in the Supplementary Information (SI). 
The above system of equations \eqref{free_energy_functional}--\eqref{eq:fmod} is numerically integrated using the finite element method (FEM) implemented in the software package \texttt{oomph-lib} \cite{Heil2006,Heil_oomph-lib_2022}. This package enables efficient direct time simulations, incorporating adaptive time stepping based on a second-order backward differentiation method. The software's spatial adaptivity makes it suitable for handling large-scale systems and sharp geometric features like wetting ridges. Simulations of Eqs.~\eqref{free_energy_functional}-\eqref{eq:fmod} were performed on one-dimensional domains of size $L=2000$ with Neumann boundary conditions.

\begin{figure}[hbt]
\includegraphics[width=\hsize]{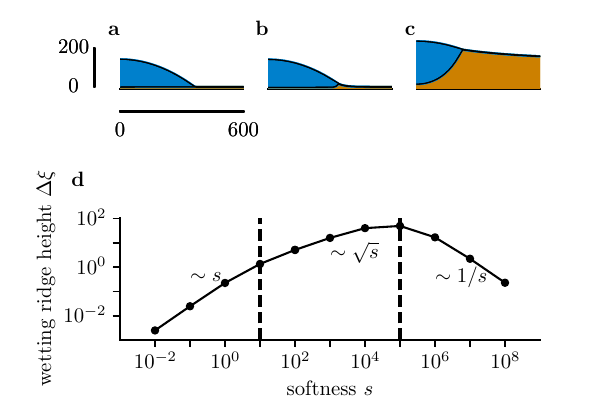}
\caption{Equilibrium shapes of passive biofilms with a fixed biomass $\int \Psi\, dx$ on soft substrates. (a-c) Typical equilibrium droplet shapes and substrate deformations in (a)  the rigid limit $s<10$,  (b) the intermediate softness regime $10<s<10^5$ regime and (c) the very soft (liquid) limit $s>10^5$. Panel (d) shows the scaling behaviour of the wetting ridge height $\Delta \xi$ that distinguishes the three regimes. The two vertical dashed lines in (d) indicate the transitions from the rigid to the soft regime and from the soft to the liquid regime.}
\label{fig:passive} 
\end{figure}

\begin{figure*}[hbt]
	\centering
	\includegraphics[width=\hsize]{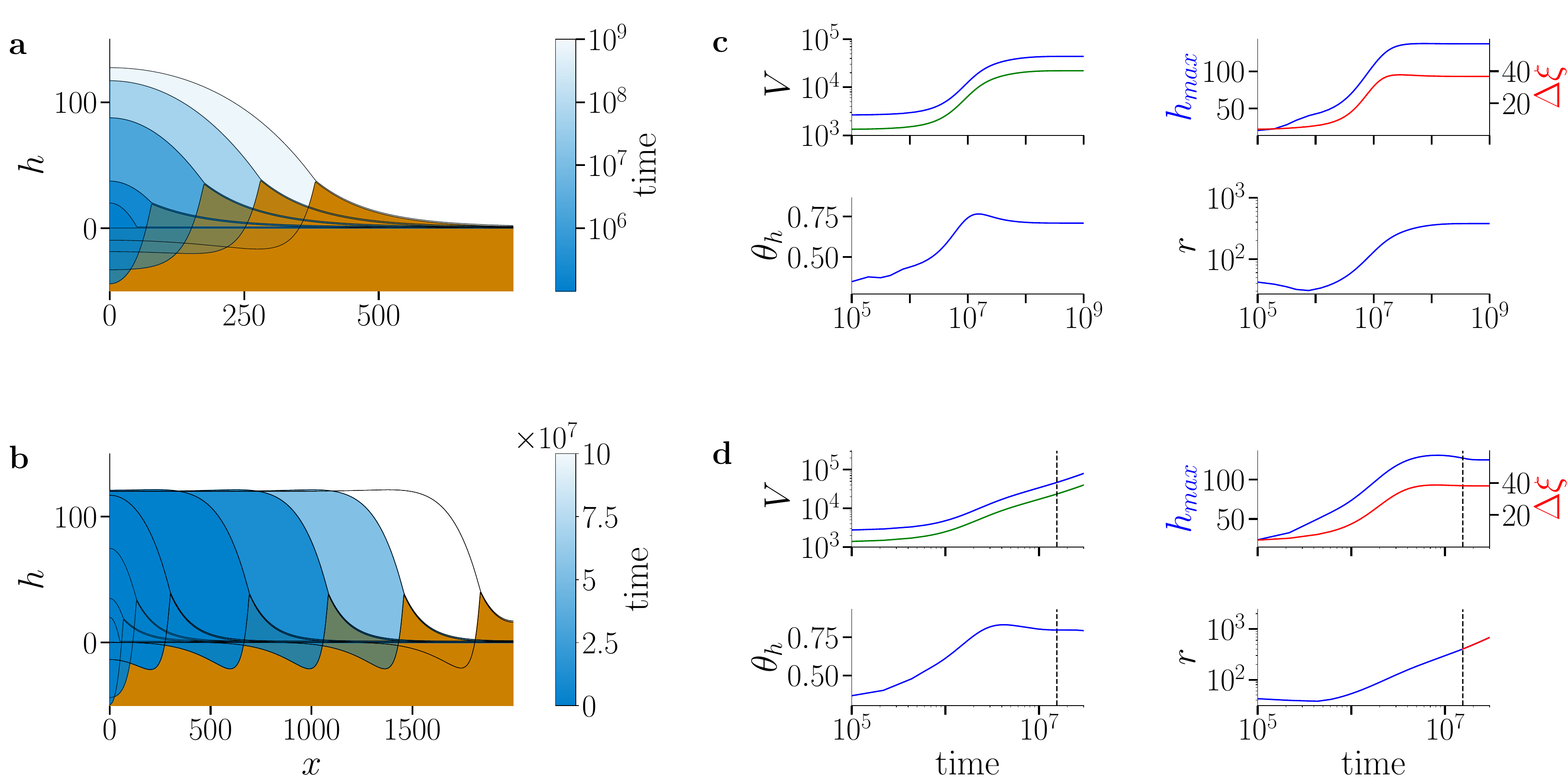}
	\caption{
          Spreading of active biofilms on soft substrates. (a,b) Snapshots of biofilm profiles (blue shading) and substrate deformation profiles (brown shading) at (a) low biomass growth rate constant ($g=1.3 \cdot 10^{-6}$) and (b) high biomass growth rate constant ($g=5.3 \cdot 10^{-6}$). The colour gradient in the biofilm profile indicates the progression of time: darker colours represent initial evolution steps and lighter colours indicate later stages. The profiles correspond to times $t \in \{0, 5.49 \cdot 10^6, 1.37 \cdot 10^7, 3.16 \cdot 10^7, 4.44 \cdot 10^8\}$ in (a) and $t \in \{0, 1.3 \cdot 10^6, 3.1\cdot 10^6, 1 \cdot 10^7, 3 \cdot 10^7, 5 \cdot 10^7, 7 \cdot 10^7, 9 \cdot 10^7\}$ in (b). Note that the colour scale is logarithmic in (a) and linear in (b). In (a) the film, biomass and substrate deformation profile evolve towards a stationary state (arrest of spreading). In (b), after an initial swelling phase, the biofilm edge advances at constant shape and speed (continuous spreading). (c,d) Time evolution of biofilm and biomass volume ($V$, blue and green, respectively), maximal biofilm ($h_{max}$,blue) and wetting ridge ($\Delta\xi$, red) heights, macroscopic contact angle $\theta_h$ (for details of measurement see SI) and biofilm extension ($r$).  The dashed vertical line in (d) denotes the onset of the continuous spreading regime. The substrate softness is $s= 10^{4}$ and remaining parameters are as given in Table~\ref{tab::listed-all-parameter} in the SI.}
	\label{fig:soft} 
\end{figure*}

\section{Passive behaviour of a biofilm on a soft elastic substrate}
\label{sec:passive-results}

In a first set of simulations, we characterise the passive spreading dynamics of biofilm droplets of fixed constant biomass $\int\Psi\,dx$, i.e.\  without biomass growth (\( g=0 \)), depending on the substrate softness $s$. In this scenario droplets may exchange solvent with the substrate which acts as an osmotic chemostat. The equilibrium state is characterised by a vanishing of all fluxes, a stationary droplet shape and a stationary substrate deformation.

 We initialised all droplets with the identical parabolic film profile $h_0(x)$ with the Young-Dupré contact angle \cite{Henkel2021,Thiele2018a} on an undeformed substrate
 and an initial profile for the biomass height $\Psi_0(x)=h_0(x)\phi_\text{eq}$ (for more details refer to the SI). Droplets then evolve towards their equilibrium shapes according to Eqs.~\eqref{free_energy_functional}-\eqref{eq:fmod}. 
 Fig.~\ref{fig:passive}~(a-c) illustrates the resulting equilibrium drop shapes and substrate deformations as a function of substrate softness $s$.
 On a rigid substrate the droplet does not deform the substrate (Fig.~\ref{fig:passive}~(a)). 
As softness increases, wetting and capillary forces increasingly deform the substrate, impacting in turn the drop shape. The first apparent signature of elastocapillary forces is the deformation of the substrate in the contact line region: with increasing softness a so-called wetting ridge is increasingly pulled out of the substrate in the vertical direction (Fig.~\ref{fig:passive}~(b)). For very soft substrates, the distinguished wetting ridge shrinks again as the droplet rather sinks into the substrate forming a liquid lens (Fig.~\ref{fig:passive}~(c)). Note that the resulting equilibrium droplet volume $V$ only depends very weakly on the substrate softness $s$ (see Fig.~\ref{fig:passive_volume} in the SI).

 As characterised for drops of simple passive liquids on soft elastic substrates \cite{Henkel2021, Andreotti2020}, three different regimes of elastocapillarity can be identified.
The three regimes can formally be distinguished by the scaling of the height of the wetting ridge\footnote{Here we define the wetting ridge height as the difference $\Delta \xi_\text{max}=\xi_\text{max}-\xi(L)$, where $\xi_\text{max}$ denotes the maximal vertical substrate displacement in the contact line region and $\xi(L)$ indicates the substrate displacement in the adhesion layer far away from the biofilm edge.} $\Delta \xi_\text{max}$ with the substrate softness \cite{Henkel2021} (Fig.~\ref{fig:passive}~(d)). Briefly, we identify a rigid regime ($s<1$), where the drop barely 'feels' the substrate softness and the wetting ridge height scales linearly with softness, i.e.\ $\Delta \xi_\text{max}\sim s^{1}$, but remains negligible compared to the droplet size (Fig.~\ref{fig:passive}~(a)). In the soft regime ($10<s<10^5$) a wetting ridge with a height smaller but comparable to the drop size is pulled out (Fig.~\ref{fig:passive}~(b)). Its height scales with softness as $\Delta \xi_\text{max}\sim s^{1/2}$. In the very soft regime, where the droplet sinks into the substrate the wetting ridge height decreases as $\Delta \xi_\text{max}\sim s^{-1}$ (Fig.~\ref{fig:passive}~(c)).
The transitions between the three regimes are associated with the interplay of three length scales, namely, the elastocapillary length $\mathcal{L}_{ec}$, the mesoscopic scale $\mathcal{L}$, i.e.\ the interface width which is governed by capillarity and wetting forces, and the macroscopic scale, i.e.\ the typical drop size $r$ \cite{lubbers2014drops,Andreotti2020, Henkel2021}. In the rigid regime $\mathcal{L}_{ec}/\mathcal{L}=\sqrt{s}\le 1$, elastic forces completely dominate, such that the substrate barely deforms, and the droplet shape is governed by the Young-Dupré equation. In the very soft regime $\mathcal{L}_{ec}>r$, capillarity and wetting forces dominate while elastic forces can be neglected. In the examples of Fig.~\ref{fig:passive}, the droplet size $r$ is of order $100\mathcal{L}$, situating the transition from the soft to the very soft regime at $s\approx 10^4$, which well agrees with our simulations. Having studied the case of passive drops as a reference case, next we consider the bioactive case.

\begin{figure*}[h!]
\centering
\includegraphics[width=0.8\hsize]{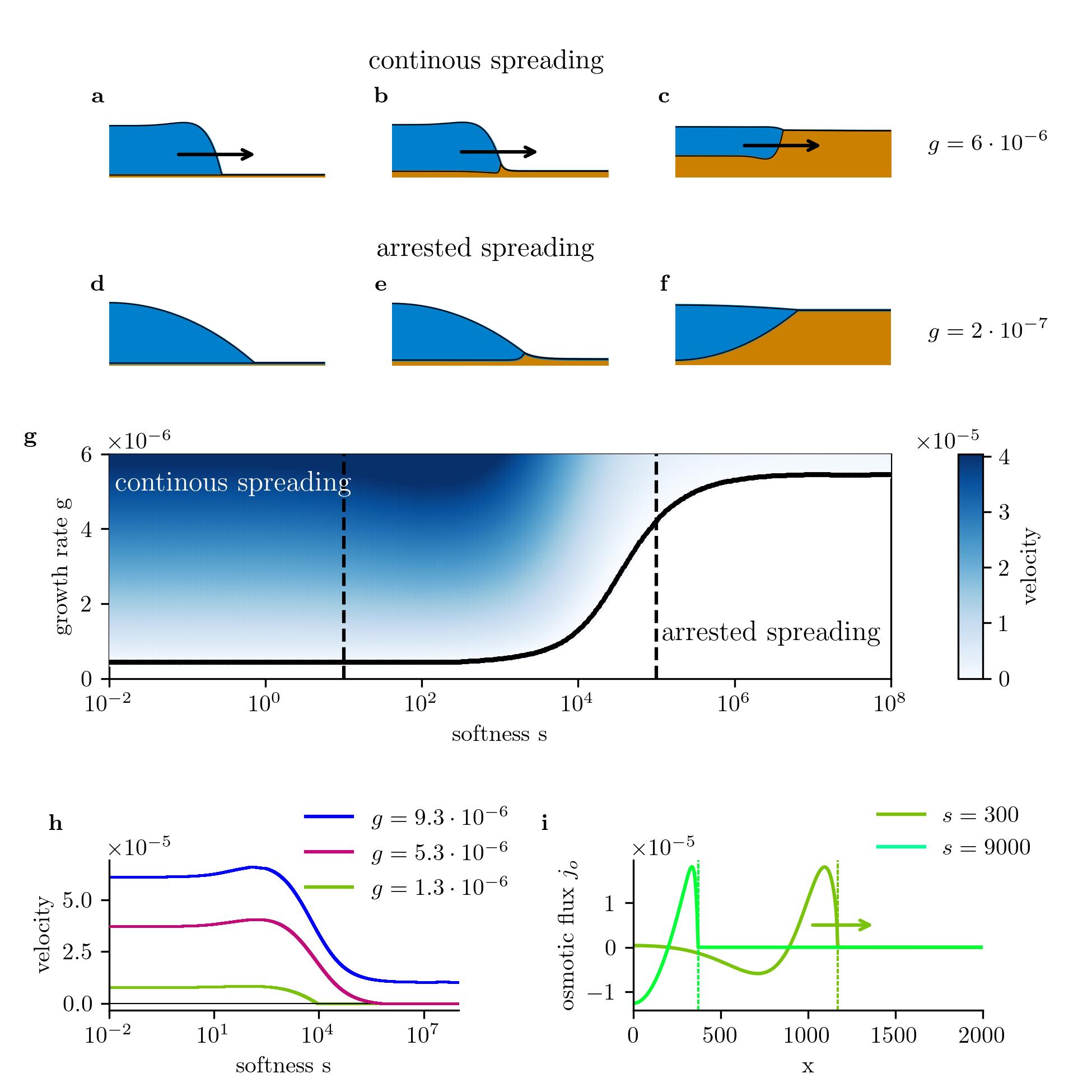}
\caption{Dynamical phase behaviour of biofilm spreading in dependence of biomass growth and substrate softness. (a-c) Drop and substrate deformation profiles for continuously spreading biofilms at a biomass growth rate constant $g=6 \cdot 10^{-6}$. The softness is (a) $s=10^{-2}$ (rigid case), (b) $s=10^{3}$ (soft case), and (c) $s=10^{7}$ (very soft case). (d-f) Drop and substrate deformation profiles for arrested spreading showing the resulting stationary biofilms at a biomass growth rate constant $g=2 \cdot 10^{-7}$. The softness is (d) $s=10^{-2}$,  (e) $s=10^{3}$, and  (f) $s=10^{7}$.
(g) Morphological phase diagram in the plane spanned by substrate softness $s$ and the biomass growth rate constant $g$. The lateral spreading velocity in the continuously spreading regime is indicated by the blue shading (see colour bar in the right). In the white region the biofilm spreading is arrested.
(h) Examples of the dependence of the lateral spreading speed on softness for various biomass growth rate constants $g$ as indicated in the legend.
(i) Examples of osmotic solvent flux profiles $j_o(x)$ for a continuously spreading biofilm at $s=300$ and an arrested biofilm at $s=9000$ at fixed $g=1.3 \cdot 10^{-6}$ (corresponding to the green curve in panel (h)). The vertical dotted lines indicate the position of the wetting ridge.
}
\label{s-g-scan} 
\end{figure*} 

\section{Spreading of active biofilms on soft substrates}
\label{sec:active-results}
Here we focus on an investigation of  the spreading behaviour of bioactive films on soft and very soft substrates (i.e.\ $s>10$, $g>0$). Note that the osmotic spreading of biofilms on rigid solid substrates has been extensively characterised by \citet{Trinschek2016,Trinschek2017}. Simulations of biofilm growth are started from a small biofilm droplet (parabolic film profile $h_0$ with biomass profile $\Psi_0=h_0\phi_\text{eq}$, see SI for details), which then develops according to Eqs.~\eqref{free_energy_functional}-\eqref{eq:fmod} with a biomass growth rate constant $g>0$. 
Fig.~\ref{fig:soft} exemplarily characterises the spreading behaviour at fixed substrate softness $s=10^{4}$ for two different biomass growth rates. 

At low biomass growth rate (\(g = 1.3 \cdot 10^{-6}\), Fig.~\ref{fig:soft}~(a)) the biofilm swells until reaching a stationary state where drop, biomass and substrate deformation profiles do not change any more. This parallels the state of arrested growth already described in \cite{Trinschek2017}, with the addition, that the substrate is deformed and a stationary wetting ridge develops at the biofilm edge. 
Strikingly, at an increased biomass growth rate constant (\(g = 5.3 \cdot 10^{-6}\), Fig.~\ref{fig:soft}~(b)), the spreading behaviour dramatically changes. After a transient phase of combined vertical and horizontal swelling, similar to the transient in the arrested case, the biofilm enters a continuous lateral spreading regime. Then the biofilm edge advances with a constant shape and speed. This behaviour mirrors the continuous growth state described in \cite{Trinschek2017} with the additional feature that the biofilm edge deforms the substrate and a wetting ridge continuously advances with the biofilm edge. 
Figs.~\ref{fig:soft}~(c,d) further characterise the arrested and continuous spreading regimes at long times in terms of the evolution of the biofilm volume and total biomass ($V$, blue and green curves, respectively), the maximal film height $h_\text{max}$ and wetting ridge height $\Delta\xi$, the contact angle of the biofilm edge $\theta_h$, and the lateral biofilm extension $r$. The short-time behaviour is shown in more detail in Fig.~\ref{fig:short:time:evolution} in the SI.
In the arrested case, at large times, the film volume, biomass and lateral biofilm extension reach plateau values whereas in the continuous growth case they increase linearly with time. In both cases, the maximal value of the film height reaches a constant value close to the imposed maximal film height $h^\ast$, where the net biomass increase stops [Eq.~\eqref{eq:growth-term}]. In the case of continuous spreading the growth restriction leads to a pronounced pancake-like shape of the biofilm, whereas in the arrested case, the biofilm adopts a spherical cap-like shape.
Furthermore, in both cases, the wetting ridge height converges to the value found in the passive limiting case (for a detailed analysis see Fig.~\ref{s-g-scan}). Similarly, the contact angle is evolving in both cases in a similar manner. 

We then investigated the robustness of the above described transition between arrested and continuous biofilm growth by varying the substrate softness $s$.
Figs.~\ref{s-g-scan}~(a) to \ref{s-g-scan}~(f) show exemplary biofilm profiles for various growth rate constants and substrate softnesses $s$ while Fig.~\ref{s-g-scan}~(g) presents a nonequilibrium phase diagram that gives the spreading speed in the parameter plane spanned by the biomass-growth rate $g$ and the substrate softness $s$. 
In general, biofilm spreading is favoured on rigid substrates. In contrast, soft substrates are unfavourable for biofilm spreading and may completely stop biofilm evolution. At very low growth rates, continuous spreading can not be achieved, independent of the substrate softness. 
Importantly, in the intermediate regime, spanning one order of magnitude for the growth rate constant g ($5\cdot 10^{-7}<g<5 \cdot 10^{-6}$), the biofilm spreading velocity is controlled by substrate softness in a parameter region spanning nearly 3 orders of magnitude, i.e.\ $10^2<s<10^5$ (Figs.~\ref{s-g-scan}~(g) and (h)).
At large growth rates ($g>5\cdot 10^{-6}$, continuous spreading occurs across all three softness regimes such that even on very soft substrates spreading is continuous, albeit with a low velocity (blue curve in Fig.~\ref{s-g-scan}~(h)). 
Note, that at high growth rates the velocity does not drop monotonically with increasing substrate softness, but slightly increases with increasing softness before dropping sharply (blue and red curves in Fig.~\ref{s-g-scan}~(h)).
For rigid substrates \citet{Trinschek2017} find that the transition from arrested to continuous spreading occurs when reducing the influence of surface forces as compared to entropic forces. Here, a similar transition is observed when varying the substrate elasticity. This shows that substrate softness represents another important passive material property, which fundamentally alters biofilm spreading behaviour.

To get a better understanding of the physical mechanism underlying the arrest of spreading on very soft substrates we next investigate the hypothesis, that the osmotic solvent exchange between the substrate and the biofilm is altered in spreading biofilms as compared to arrested biofilms.
Fig.~\ref{s-g-scan}i shows exemplary spatial profiles of the osmotic exchange flux $j_o(x)$ between the substrate and the biofilm for a spreading (dark green) and an arrested (light green) biofilm. 
In the spreading biofilm, the flux vanishes in the central region of the biofilm ($x\approx 0$). Close to  the biofilm edge, where biomass is produced, the osmotic flux is directed from the substrate into the biofilm. Interestingly, in a region between the biofilm edge and the biofilm centre, the osmotic flux is directed from the biofilm into the substrate. However, the osmotic outflow in this region is weak compared to the massive osmotic influx at the advancing biofilm edge.
In the arrested case, osmotic fluxes are still present. Although the spatial profiles (film height $h$, biomass $\Psi$, substrate deformation $\xi$) are stationary, and conserved and nonconserved fluxes are balanced, they do not vanish. This is a signature of active, i.e., out-of-equilibrium, behaviour and is a direct results of the gradient dynamics structure being broken by biomass growth $j_g$ in Eq.~\eqref{gradient_dynamics_Psi_t}. 
The osmotic flux is directed into the substrate at the centre of the biofilm and into the biofilm close to the edge of the biofilm, where it drops to zero as the wetting ridge is approached.
From the inspection of the osmotic flux profiles on the macroscale it does not yet become clear, what feature causes the spreading arrest, since the observed osmotic influx close to the biofilm edge should favour spreading in both cases.

\begin{figure}[h!]
\includegraphics[width=\hsize]{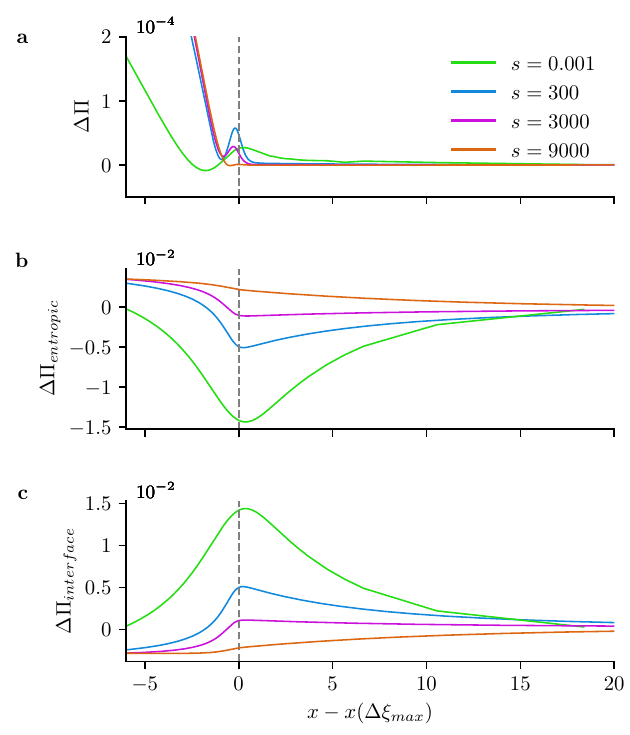}
\caption{Role of the osmotic flux between the substrate and the biofilm in spreading on soft substrates. (a) Osmotic pressure difference $\Delta\Pi$ between biofilm and substrate in the contact line region for various agar softnesses $s$ as indicated in the legend. The growth rate constant is $g=1.3 \cdot 10^{-6}$. At a softness of $s=9000$ the contact line is stationary. All other curves ($s<9000$) belong to spreading biofilm fronts. 
(b) Entropic  $\Delta\Pi_\text{entropic}$, see Eq.~\ref{eq:pi-entropic},  and (c) interfacial $\Delta\Pi_\text{interface}$, see Eq.~\ref{eq:pi-interfacial}, contributions to the osmotic pressure difference. The legends are as in (a). The grey dashed lines in (a-c) indicate the location of the wetting ridge for all profiles. Remaining parameters are given in Table \ref{tab::listed-all-parameter} in the SI.}
\label{fig:osmotic:pressure}
\end{figure}

However, a close inspection of the osmotic pressure profile in the biofilm on the mesoscopic scale in the contact line region reveals important differences between arrested and spreading biofilms.
Fig. \ref{fig:osmotic:pressure}a shows the spatial profile of the osmotic pressure difference $\Delta \Pi$ between the biofilm and the substrate in the contact line region for spreading and arrested biofilms. All profiles were laterally shifted such that the wetting ridge is located at $x=0$.
The osmotic pressure difference $\Delta \Pi$ for the advancing biofilm fronts ($s<9000$) is positive and shows a local maximum in the contact line region $x\approx 0$ indicating solvent influx into the biofilm. However, for the arrested biofilm front at $s=9000$ the osmotic pressure difference $\Delta \Pi\approx 0$. The osmotic pressure difference contains two contributions: one results from the entropy of mixing of biomass $\Delta \Pi_\text{entropic}$ and another one results from interfacial forces (capillarity and wettability) $\Delta\Pi_\text{interface}$. Using expression \eqref{eq:Pi_s} for the constant osmotic pressure $\Pi_s=-\frac{k_BT}{a^3} \ln{(1-\phi_\text{eq})}$ in the substrate, the two contributions can be expressed as (see Eqs.~\eqref{eq:osm:flux} and \eqref{def:Pi})
\begin{align}
 \label{eq:pi-entropic} \Delta\Pi_\text{entropic} & =  \frac{k_BT}{a^3}\left[\ln{(1-\phi_\text{eq})}-\ln{(1-\phi)}\right], \\
 \label{eq:pi-interfacial} \Delta\Pi_\text{interface} & =  \gamma_h\Delta(h+\xi)   -   \frac{\partial f_\text{w}}{\partial h}
 \end{align}
The decomposition of the osmotic pressure difference $\Delta\Pi$ into entropic and interfacial contributions (Fig.~\ref{fig:osmotic:pressure}~(b,c)) reveals that for advancing biofilm edges ($s\le 3000$) at the contact line the entropic contribution favours osmotic outflux ($\Delta\Pi_\text{entropic}<0$) since the biomass concentration is below the equilibrium value ($\phi<\phi_\text{eq}$). However, the interfacial contribution favours osmotic influx ($\Delta\Pi_\text{interface}>0$). The resulting net flux is small ($10^{-4}$, compared to the magnitude of $10^{-2}$ for each individual flux) but positive, driving the advancement of the biofilm edge. 
In contrast, for arrested fronts ($s=9000$ in Fig.~\ref{fig:osmotic:pressure}~(b,c)) the biomass concentration in the biofilm $\phi$ favours solvent influx, whereas the interfacial forces are inducing a negative osmotic pressure difference, such that outflux is favoured. Importantly, the net solvent flux between substrate and biofilm vanishes at the contact line. Note, that a fully developed advancing biofilm front (that has developed on a rigid substrate) will initially recede until the biofilm adopts the stationary droplet shape on a soft substrate. 

In conclusion, arrested and spreading case differ qualitatively w.r.t.\ the forces which drive osmotic fluxes. The transition between the two regimes results from a subtle equilibration of these forces in the contact line region. It emerges that direction and magnitude of osmotic  solvent fluxes are decisive for biofilm spreading. Viscoelastic braking observed in droplet spreading of passive fluids is certainly present in biofilms as well, but not the major determinant for a decreased spreading speed on soft to very soft substrates.

\section{Comparison with experiments}
\label{sec:comp}

\begin{figure*}[h!]
\includegraphics[width=\hsize]{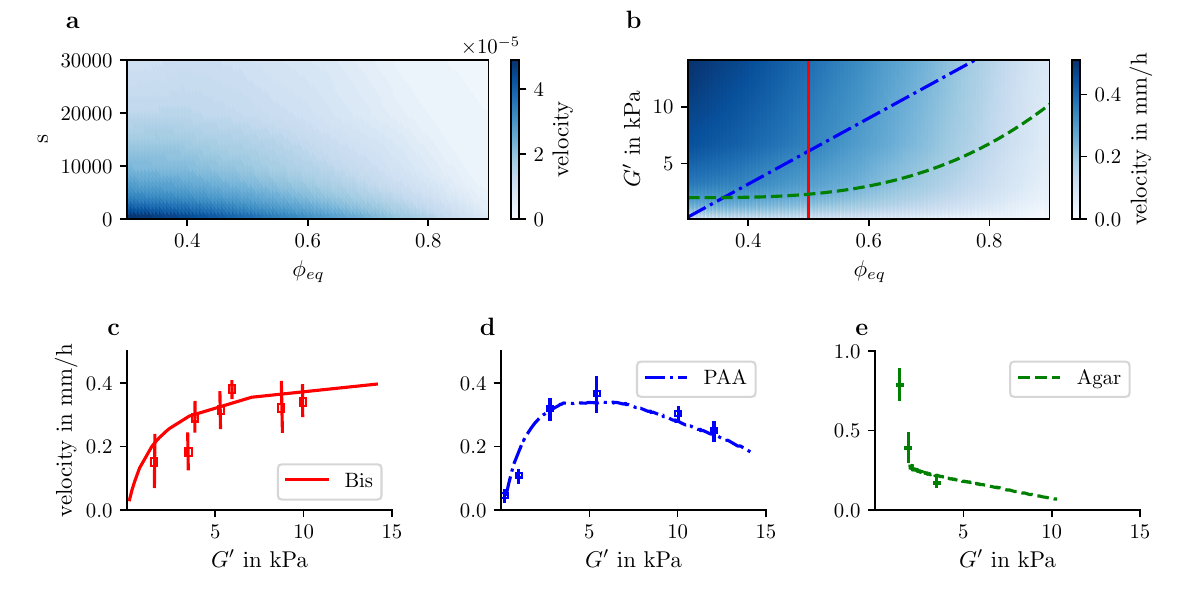}
\caption{(a,b) Phase diagrams characterising the biofilm spreading velocity in the plane spanned by the equilibrium biomass concentration \( \phi_{\text{eq}} \) and (a) on the one hand the softness $s$ and (b) on the other hand the elastic shear modulus \( G' \). Thereby, the colour (scales given on the right of each panel) indicates the lateral (adimensional (a), dimensional (b)) spreading velocity within the continuous spreading regime.
(c-e) Biofilm spreading velocity depending on the substrate shear modulus $G'$ for three exemplary scenarios $G'(\phi_\text{eq})$ as indicated by matching coloured lines in (b), see main text. Lines represent theoretically calculated spreading velocities; symbols correspond to the experimental data of \citet{Asp2022} on three different substrate architectures, namely,
(c) PAA substrate with fixed PAA concentration and varying cross-linker (bis-acrylamide) concentration,
(d) PAA substrate with varying PAA concentration and constant cross-linker (bis-acrylamide) concentration, and
(e) Agar substrate with varying agar concentration. 
The (nondimensional) growth rate constant is $g = 5.3 \cdot 10^{-6}$.
The dimensional velocity and the shear modulus are determined as described in the main text.}
\label{fig:experimental}    
\end{figure*}

Up to now, we have treated the osmotic pressure in the substrate (determined by the hydrogel concentration $\phi_\text{eq}$) and the substrate softness $s$ as independent parameters, and have obtained results as discussed at Figs.~\ref{fig:soft} and~\ref{s-g-scan} at fixed $\phi_\text{eq}$. 
This ideal assumption of independent parameters only approximately holds for specific experimental protocols of substrate fabrication. Typically, the hydrogel substrate rigidity is varied, e.g. by increasing the cross-linker concentration at constant hydrogel concentration or by changing the hydrogel concentration at constant cross-linker concentration. 
When the latter method is used both parameters, osmotic pressure and elastic modulus of the substrate, are impacted in a way specific for each hydrogel.
To be able to discuss such scenario, we study the influence of the osmotic pressure of the substrate (via the equilibrium biomass concentration $\phi_{eq}$ \cite{Kochanowski2024,Charlton2022}). Fig.~\ref{fig:experimental}~(a) shows the resulting morphological phase diagram in the plane spanned by $s$ and $\phi_{eq}$ at fixed biomass growth rate $g=5.3\cdot 10^{-6}$. Generally speaking, rigid substrates and a low substrate osmotic pressure favour a rapid biofilm expansion. An increase in the biofilm osmotic pressure reduces the propagation speed of the biofilm edge, since a larger amount of biomass has to be produced in the biofilm to draw water from the substrate into the biofilm.

To establish at least a semi-quantitive link between theory and experiment based on the obtained phase diagram, we introduce (i) appropriate scales to relate the nondimensional softness and the elastic modulus as well as the nondimensional spreading speed to the dimensional one and (ii) we relate substrate osmotic pressure and elastic modulus for a given experimentally employed substrate architecture. 

Regarding (i), the shear modulus $G'$ scales as $G'\sim \kappa_v d$, with $\kappa_v$ denoting the model elastic constant of the substrate and $d$ denoting the substrate thickness \cite{Henkel2021}. Using the definitions of softness $s$ \eqref{def:s} and elastocapillary length $\mathcal{L}_{ec}=\sqrt{\gamma_h/\kappa_v}$ we find a scaling between the linear shear modulus $G'$ and the softness $s$
\begin{equation}
G'\sim \frac{\gamma_h d}{\mathcal{L}^2 s}\,. \label{G:s}
\end{equation}
The velocity scale is given by $\mathcal{L}/\mathcal{T}$ with the length scale $\mathcal{L}$ and time scale $\mathcal{T}$ defined in \eqref{eq:L} and \eqref{eq:T}.

With this scaling we now compare our results with the experimental observations reported by \citet{Asp2022} (see their Fig.~4). They investigate biofilm spreading on hydrogel substrates of various chemical compositions and architectures (agar substrate, bis-acrylamide (Bis) cross-linked polyacrylamide (PAA)). Thereby, they influence the elastic modulus by varying either the polymer concentration (agar, PAA) or the cross-linker concentration (Bis). Osmotic spreading on agar substrates with varying concentrations has also been studied, e.g.\ in Refs.~\cite{Kochanowski2024, Ziege2021} with similar observed trends as reported in \citet{Asp2022}.

To proceed we assume a (nondimensional) growth rate constant $g=5.3\cdot 10^{-6}$ and visually compare our Fig.~\ref{s-g-scan}~(h) with Fig.~4 of~\citet{Asp2022}. In this way, we are able to establish the relation $s G'=\gamma_h d/\mathcal{L}^2=2\cdot 10^7$\,Pa between $s$ and $G'$ and to determine the velocity scale to be $\mathcal{L}/\mathcal{T}=2.8\,$mm s$^{-1}$. Using the gel thickness $d=1\,$mm \cite{Asp2022} and the surface tension and viscosity of water ($\gamma_h=70 \cdot 10^{-3}\,$N\,m$^{-1}$, $\eta_0=10^{-3}$\,Pa s), we obtain the lateral length scale $\mathcal{L}=1.9\,\mu$m, the vertical length scale $\mathcal{H}=0.1\,\mu$m, and the time scale $\mathcal{T}=7\,\mathrm{s}^{-4}$. These scales are consistent with a Hamaker constant $A=10^{-18}$\,N\,m and a dimensional biomass (extracellular matrix, bacteria) growth rate constant $g=0.5$\,min$^{-1}$.
The scaling procedure above results in Fig.~\ref{fig:experimental}~(b) that gives the rescaled representation of the biofilm spreading speed in the plane spanned by the equilibrium biomass concentration $\phi_{\text{eq}}$ and the elastic shear modulus $G'$. Also here we see that more rigid substrates with low osmotic pressure favour a rapid biofilm expansion.

Finally, we come back to point (ii) from above, regarding the relation between parameters that we have treated as independent up to now. Identifying the experimental parameter of hydrogel concentration with our $\phi_\text{eq}$, we consider three example scenarios for the dependence of hydrogel stiffness $G'$ on $\phi_\text{eq}$ that correspond to the experimental conditions reported by \citet{Asp2022} in their Fig.~4: In scenario one, $G'$ is increased via an increase of the cross-linker concentration at fixed PAA concentration (i.e.\ $\phi_\text{eq}$ remains constant, red solid line in  Fig.~\ref{fig:experimental}~(b) and Fig.~\ref{fig:experimental}~(c)). In scenario two, $G'\sim \phi_\text{eq}$ scales linearly with the hydrogel concentration (PAA, constant cross-linker concentration, measured by \citet{Asp2022}, blue dot-dashed line in Fig.~\ref{fig:experimental}~(b) and Fig.~\ref{fig:experimental}~(e)).  In scenario three, $G'\sim\phi^\nu_\text{eq}$ scales with an exponent $\nu\ge 2$ (agarose \cite{fujii2000scaling,Tokita1987}, green dashed line in Fig.~\ref{fig:experimental}~(b) and Fig.~\ref{fig:experimental}~(f)). 

The mentioned  Figs.~\ref{fig:experimental}~(c) to \ref{fig:experimental}~(e) compare the resulting dependence of spreading velocity on shear modulus $G'$ as experimentally measured by \citet{Asp2022} and as obtained here when following the three trajectories in the $(\phi_\text{eq}, G')$-plane (Fig.~\ref{fig:experimental}~(b)). Interestingly, depending on the hydrogel substrate and the implied specific functional dependence $G'(\phi_\text{eq})$, biofilm spreading can be enhanced with increasing substrate stiffness (Fig.~\ref{fig:experimental}~(c)), can show a nonmonotonous behaviour (Fig.~\ref{fig:experimental}~(d)), or can be slowed down with increasing substrate stiffness (Fig.~\ref{fig:experimental}~(e)). The obtained coherent explanation of seemingly contradictory results for the dependencies on $G'$ underscores the importance of understanding the relation between the hydrogel parameters elasticity and osmotic pressure.

\section{Conclusion}
\label{sec:concl}

We have demonstrated, that biofilm spreading on soft substrates is critically slowed down compared to rigid substrates and that substrate softness can even lead to the complete arrest of the biofilm edge. This result holds for the case, that mechanical (rigidity) and physico-chemical (osmotic pressure) substrate properties can be independently controlled.
Furthermore, we find, that the reduction of the biofilm spreading speed on soft substrates is not directly related to the visco-elastic braking observed for the hydrodynamic spreading of liquid droplets on soft solid substrates \cite{Karpitschka2015,Carre2001,Andreotti2020,Henkel2021} but is rather caused by a reduced osmotic influx of solvent into the biofilm at its edges, which results from the thermodynamic coupling between substrate deformation and interfacial forces.
However, for practically used hydrogel substrates often the mechanical softness and the osmotic pressure cannot be independently controlled, i.e.\ they cannot be treated as independent parameters. Typically, an increase in substrate softness, which slows down spreading, is also associated with a reduced osmotic pressure in the substrate, which in turn accelerates spreading. 
This effect is clearly illustrated by~\citet{Asp2022}: They show that the biofilm spreading velocity on hydrogels strongly depends on substrate architecture. The spreading velocity may increase or decrease with increasing substrate rigidity or may even show nonmonotonic behaviour. Coupling substrate rigidity and substrate osmotic pressure via scaling laws, our minimal theory coherently captures these seemingly contradictory experimental results and clearly underscores the importance of substrate deformations and osmotic fluxes in the dynamics of the advancing biofilm edge.

While the here proposed minimalistic model applies to biofilm growth on abiotic surfaces under air, other settings may be considered within a sligthly modified modelling framework: biofilm growth on biotic surfaces (e.g.\ airway epithelia, airway mucus; relevant for biofilm development in Cystic fibrosis patients due to \textsl{Pseudomonas aeruginosa} \cite{Matsui2006,MoreauMarquis2008}) or growth of immersed biofilms on soft abiotic and biotic surfaces \cite{DiDomenico2022,Cont2020}. In the latter case, the coupling between osmotic substrate pressure and substrate rheology is a less important feature, since osmotic fluxes between biofilm and the surrounding medium will dominate over the solvent exchange between biofilm and substrate.
While our model cannot capture important (species-dependent) biological factors of biofilm initiation and development (e.g.\ quorum sensing), it constitutes nevertheless a useful tool to understand how the physico-chemical and mechanical environmental parameters impact the dynamics of the advancing biofilm edge and may help, e.g.\ to develop alternative therapeutic strategies in a targeted manner that do not rely on antibiotic drugs.

%
\section*{Author contributions}
AP, KJ, and UT designed research. AP performed all numerical calculations. AP, KJ, and UT wrote the manuscript.
%
\section*{Conflicts of interest}
There are no conflicts to declare.
\section*{Data availability}
All data necessary to plot Figs.~\ref{fig:passive}-\ref{fig:experimental} and the supplementary Figs.~\ref{fig:passive_volume}-\ref{fig:active:ridge} are available in the zenodo repository under the  URL \url{https://zenodo.org/records/14235001}.

\section*{Acknowledgements}
This work was supported by the \textsl{Studienstiftung des Deutschen Volkes} and the Graduate School \textsl{Living Fluids} of the Franco-German University. The authors thank Sigolène Lecuyer, Delphine Débarre and Lionel Bureau for fruitful discussions.
%
%

\setcounter{section}{0}
\makeatletter 
\renewcommand{\thesection}{S\@arabic\c@section}
\makeatother
\section{Supplementary Information}
\subsection{Dimensionless model equations}

\setcounter{figure}{0}
\makeatletter 
\renewcommand{\thefigure}{S\@arabic\c@figure}
\makeatother

\setcounter{table}{0}
\makeatletter 
\renewcommand{\thetable}{S\@arabic\c@table}
\makeatother

Starting out from Eqs.~\eqref{eq:phi}--\eqref{eq:fmod} in section \ref{sec:model} we derive here the evolution equations for $h$, $\Psi$ and $\xi$ in their dimensionless form.
In a first step we rewrite Eqs.~\eqref{gradient_dynamics_h_t}--\eqref{gradient_dynamics_xi_t} in a more convenient form
\begin{align}
	\partial_{t} h &= - \nabla .\textbf{j}_{\text{conv}} + j_o \\
	\partial_{t}\Psi &= - \nabla . (\phi \textbf{j}_{\text{conv}} + \textbf{j}_{\text{diff}}) + j_g \\
	\partial_{t} \xi &=	j_s\,.
\end{align} 
with the convective flux $\textbf{j}_\text{conv}$
\begin{equation}
\textbf{j}_\text{conv} = -\frac{h^3}{3\eta}\nabla\left[\partial_h f_w-\gamma_h\Delta(h+\xi)\right]\,,
\end{equation}
the diffusive flux $\textbf{j}_\text{diff}$
\begin{equation}
\textbf{j}_\text{diff}=-Dh\phi\nabla \partial_\phi f_m\,,
\end{equation}
and the substrate motion
\begin{equation}
j_s=-\frac{1}{\zeta}\left[\kappa_v\xi-\gamma_h\Delta(h+\xi)-\gamma_\xi\Delta\xi
\right]\,.
\end{equation}
The osmotic flux $j_o$ and the biomass growth $j_g$ are defined in Eqs.~\eqref{eq:osm:flux}--\eqref{eq:fmod} in section \ref{sec:model}.

Introducing the vertical and horizontal length scales, $\mathcal{H}$ and $\mathcal{L}$, and time scale $\mathcal{T}$ as defined in \eqref{eq:H}, \eqref{eq:L} and \eqref{eq:T} we find the evolution equations for the adimensional quantities $\hat{h}=h/\mathcal{H}$, $\hat{\Psi}=\Psi/\mathcal{H}$, $\hat{\xi}=\xi/\mathcal{H}$
\begin{align}
\partial_{\hat{t}} \hat{h} & = -\hat{\nabla}.\hat{\textbf{j}}_\text{conv}+\hat{j}_o\\
\partial_{\hat{t}} \hat{\Psi} &=  - \hat{\nabla} . (\phi \hat{\textbf{j}}_{\text{conv}} + \hat{\textbf{j}}_{\text{diff}}) + \hat{j}_g \\
\partial_{\hat{t}} \hat{\xi} &=	\hat{j}_s\,,
\end{align}
where $\hat{}$ indicates adimensional quantities, e.g. $\hat{\nabla}=\mathcal{L}\nabla$.
The dimensionless convective flux $\hat{\textbf{j}}_\text{conv}$ takes the form
\begin{equation}
\hat{\textbf{j}}_\text{conv}=-\frac{\hat{h}^3}{\hat{\eta}}\hat{\nabla}\left[\frac{1}{\hat{h}^3}-\frac{1}{\hat{h}^6}-\hat{\Delta}(\hat{h}+\hat{\xi})\right]\,.
\end{equation}
with $\hat{\eta}=1-\phi+\mu\phi$.
The dimensionless diffusive flux $\hat{\textbf{j}}_\text{diff}$ takes the form
\begin{equation}
\hat{\textbf{j}}_\text{diff}=\frac{\hat{D}\hat{h}\phi}{\hat{\eta}}\hat{\nabla}\ln\left(\frac{\phi}{1-\phi}\right)\,.
\end{equation}

The unconserved fluxes are given in their dimensionless form by
\begin{align}
\hat{j}_o & = \hat{Q}_o\left[\hat{\Delta}(\hat{h}+\hat{\xi})-\left(\frac{1}{\hat{h}^3}-\frac{1}{\hat{h}^6}\right)-\frac{1}{W_m}\ln\left(\frac{1-\phi}{1-\phi_{eq}}\right)\right]\\
\hat{j}_g & = \hat{g}\hat{h}\phi\left(1-\phi\right)\left(1-\frac{\hat{h}\phi}{\hat{h}^\ast\phi_{eq}}\right)\left(1-\frac{\hat{h}_u\phi_{eq}}{\hat{h}\phi}\right)\left[1-e^{\left(\phi_{eq}-\hat{h}\phi\right)}\right]\\
\hat{j}_s & =\frac{\sqrt{s}}{\tau}\left[\hat{\Delta}(\hat{h}+\hat{\xi})+\sigma\hat{\Delta}\hat{\xi}-\frac{1}{s}\hat{\xi}\right]\,. 
\end{align}
Note, that we have assumed that the friction constant of the substrate $\zeta$ is related to the bulk viscosity of the substrate $\eta_s$ as $\frac{1}{\zeta}=\sqrt{\frac{\gamma_h}{\kappa_v}}\frac{1}{\eta_s}$ \cite{Henkel2022}. 
The dimensionless parameters are 
\begin{align}
\hat{D} & =\frac{k_BTh_a}{2\pi A a}\\
W_m & =\frac{Aa^3}{k_BT h_a^3}\\
\hat{Q}_o & = \frac{\mathcal{T}\gamma_hQ_o}{\mathcal{L}^2} =\frac{3Q_o\gamma_h\eta_0 h_a}{A}\\
\hat{g} & = \mathcal{T}g\\
\tau & = \frac{\eta_s\mathcal{L}}{\gamma_h\mathcal{T}} =\frac{\eta_s}{3\eta_0}\left(\frac{A}{\gamma_h h_a^2}\right)^{3/2}\\
\sigma & = \frac{\gamma_\xi}{\gamma_h}\,.
\end{align}
The softness $s$ is defined in the main text \eqref{def:s}.
Table \ref{tab::listed-all-parameter} summarizes all adimensional parameters.

\begin{table}[h]
\small
\caption{\ Summary of all adimensional model parameters}
\label{tab::listed-all-parameter}
\begin{tabular*}{0.48\textwidth}{@{\extracolsep{\fill}}p{5cm}ll}
\hline
Parameter     & Symbol & Value \\
\hline
growth rate constant (as stated in the main text or the figure legends) & $\hat{g}$  & \\
substrate softness (as stated in the main text or the figure legends) & $s$ & \\
wetting parameter & $W_m$ & 5 \\
viscosity contrast biomass-solvent & $\mu$ & $10^{4}$ \\
diffusion constant & $\hat{D}$ & $\frac{1}{2\pi}$\\
osmotic mobility & $\hat{Q}_o$ & 0.01 \\
equilibrium biomass concentration (if not stated otherwise) & $\phi_{\text{eq}}$ & 0.5 \\
substrate relaxation time & $\tau$ & 7.6 \\ 
surface tension ratio & $\sigma$ & 0.1 \\
maximal biofilm height & $h^{*}$ & 120 \\
critical biofilm height for biomass  production & $h_u$ & 2 \\
\hline
\end{tabular*}
\end{table}

\subsection{Initial conditions}
Calculations of the biofilm evolution were started from identical initial conditions $h_0(x)$ where a small biofilm droplet of maximum height $h_\text{0,max}=20$ was approximated by a parabola of the form
\begin{equation}
h_0(x)=h_\text{0,max}-\frac{\theta_{eq}^2}{4h_\text{0,max}}x^2\,,
\end{equation}
where $\theta_{eq}$ denotes the equilibrium contact angle $\theta_{eq}=\sqrt{3/5}$ of the adimensional system \cite{Thiele2018a,Henkel2021}.
The biomass concentration is initiated as $\phi_0(x)=\phi_{eq}$, i.e. the biomass layer thickness follows the initial biofilm profile
$\Psi_0(x)=\phi_{eq}h_0(x)$. The initial state of the substrate is the reference configuration, i.e. $\xi=0$.

\subsection{Measurement of the spreading velocity and the contact angle}

In order to measure the lateral spreading velocity of the biofilm we determined first the radius $r$ of the biofilm as the position of the inflection point in the biofilm front. The velocity was then determined from the slope of the curve $r(t)$.

To calculate the contact angle \(\theta_h\) of the biofilm edge with the substrate, we fitted the droplet profile (in the arrested and spreading case) in the relevant front region with a parabola of the form \(f(x) = -a(x - b)^2 + c\). 
The assumed form of the fit function is a parabola given by
\[
f(x) = -a(x - b)^2 + c\,.
\]
The contact angle was then determined as the slope of the parabola where the biofilm front meets the substrate, i.e. $\theta_h=2\sqrt{ac}$.

\subsection{Volume dependency of a passive biofilm with a fixed biomass}

\begin{figure}[hbt]
\includegraphics[width=\hsize]{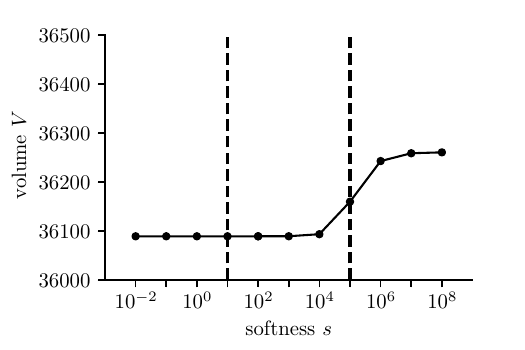}
\caption{Equilibrium states of passive biofilms with a fixed biomass $\int \Psi\, dx$ on soft substrates. Dependence of the biofilm volume $V$ on substrate softness $s$. The vertical dashed lines separate the three scaling regimes introduced in Fig.~\ref{fig:passive}. The biomass in the droplet is $\int\Psi\,dx= \phi_{eq}V_0$ with the initial volume $V_0 = \approx3.6 \cdot 10^4$ and $\phi_{eq}=0.5$. The remaining parameters are given in Table \ref{tab::listed-all-parameter}.}
\label{fig:passive_volume} 
\end{figure}

If a biofilm droplet (parabolic profile with the equilibrium contact angle $\theta_h=\sqrt{3/5}$ and with biomass concentration $\phi=\phi_{eq}$) is placed on an undeformed substrate ($\xi=0$) and the biofilm may evolve in the absence of active biomass production ($g=0$),  the droplet will eventually adopt an equilibrium shape. During the equilibration phase the biofilm will exchange solvent with the substrate, thereby changing its volume. However, the influence of the substrate softness $s$ on the equilibrium biofilm volume is negligible (Fig.~\ref{fig:passive_volume}).

\subsection{Evolution of biofilm droplets on short time scales}

\begin{figure}[hbt]
\includegraphics[width=\hsize]{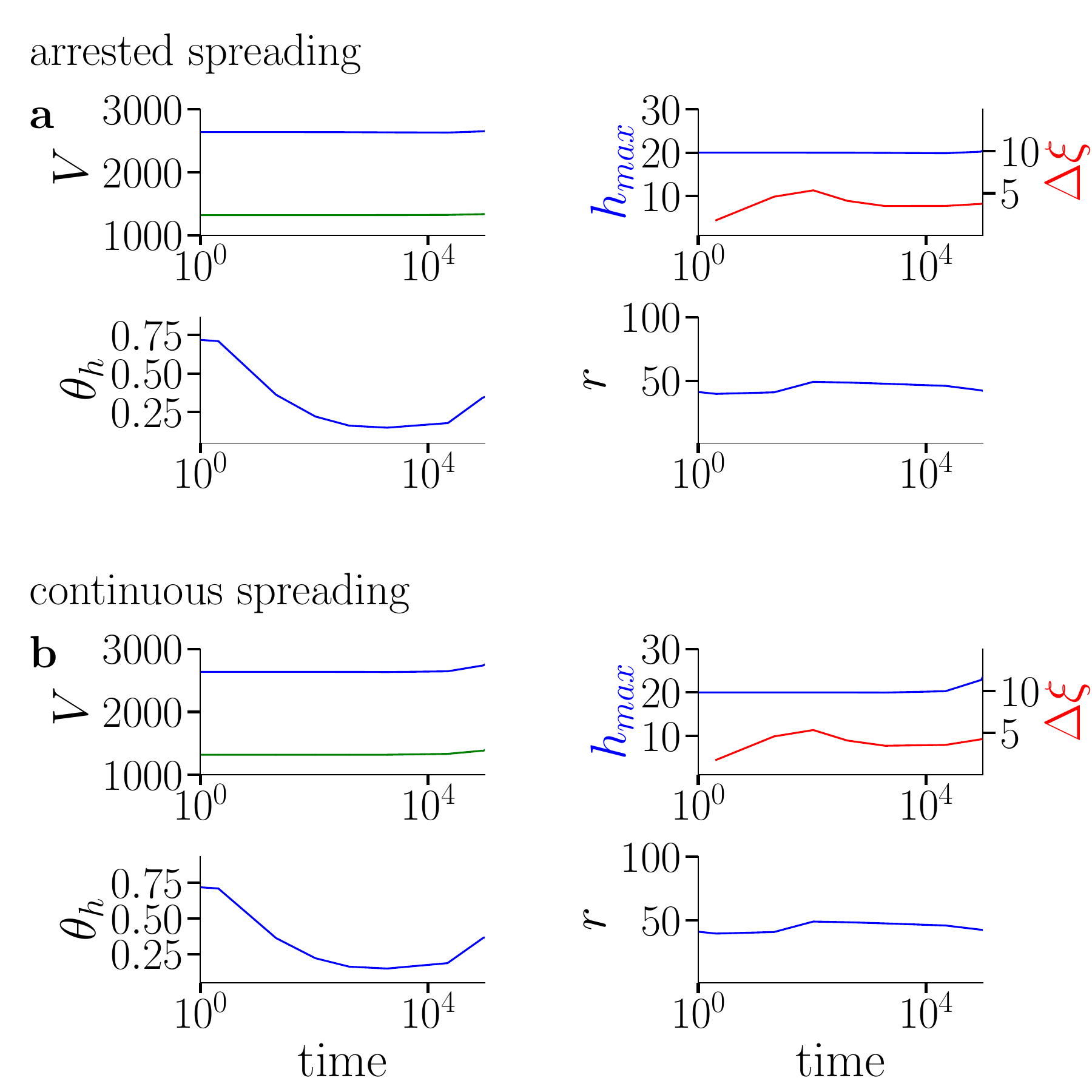}
\caption{Spreading of active biofilms on soft substrates on short time scales at low (a, $g=1.3 \cdot 10^{-6}$) and high (b, $g=5.3 \cdot 10^{-6}$) biomass growth rate constant.  Shown is the evolution at short times of biofilm and biomass volume ($V$, blue and green, respectively), maximal biofilm ($h_{max}$, blue) and wetting ridge ($\Delta\xi$, red) heights, macroscopic contact angle $\theta_h$ and biofilm extension ($r$). The substrate softness is $s= 10^{4}$ and remaining parameters are as given in Table~\ref{tab::listed-all-parameter}.}
\label{fig:short:time:evolution}
\end{figure}

Fig.~\ref{fig:short:time:evolution}~(a,b) characterises the arrested and continuous
spreading regimes at short times in terms of the evolution of
the biofilm volume and total biomass ($V$ , blue and green curves,
respectively), the maximal film height $h_{max}$ and wetting ridge
height $\Delta\xi$, the contact angle of the biofilm edge $\theta_h$, and the lateral biofilm extension $r$. The main characteristic at short times is a decrease in the contact angle, whereas all other observables do not change considerably. This behaviour holds for the arrested case (low growth rate constant) and the continuous spreading case (high growth rate constant).

\subsection{Dependency of the wetting ridge height on the substrate softness \textit{s} for an active biofilm}

\begin{figure}[hbt]
\includegraphics[width=\hsize]{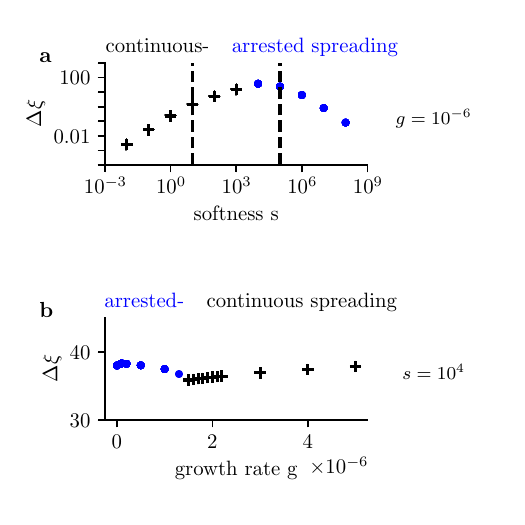}
\caption{
Dependence of the wetting ridge height on the substrate softness $s$ (a) and biomass growth rate constant $g$ (b). The vertical dashed lines separate the three scaling regimes as explained in Fig.~\ref{fig:passive}. The remaining parameters are given in Table \ref{tab::listed-all-parameter}.
}
\label{fig:active:ridge} 
\end{figure}

To get a better understanding of the substrate softness on the spreading behaviour of biofilms with biomass production ($g>0$) we measure the wetting ridge height for spreading and stationary biofilms (Fig.~\ref{fig:active:ridge}~(a)). 
The wetting ridge height is mainly determined by the substrate softness, and, even in the active case, a similar scaling behaviour as in the passive droplet case can be observed (cf.~Fig.\,\ref{fig:passive}~(d)).
Interestingly the transition between spreading and arrested regime coincides with the transition from the soft to the very soft (liquid-like) substrate, where the biofilm starts to sink into the substrate, and the height of the wetting ridge starts to decrease.
The growth rate has only a minor influence on the wetting ridge height (Fig.~\ref{fig:active:ridge}~(b)). 
However, here the transition of the arrested to the spreading regime (by increasing the growth rate constant $g$) is marked by a minimum in the height of the wetting ridge.

\end{document}